\documentclass[aps,pra,reprint,showpacs]{revtex4-1}

\usepackage{graphicx} 
\usepackage{dcolumn}
\usepackage{amsfonts}
\usepackage{amssymb}
\usepackage{amsmath}
\usepackage{bm}
\newcommand{\be}{\begin{equation}}
\newcommand{\ee}{\end{equation}}
\newcommand{\bea}{\begin{eqnarray}}
\newcommand{\eea}{\end{eqnarray}}

\begin{document}

\title{Spectroscopy of a narrow-line laser cooling transition in atomic dysprosium}

\author{Mingwu Lu}
\author{Seo Ho Youn}
\author{Benjamin L. Lev}
\affiliation{Department of Physics, University of Illinois at Urbana-Champaign, Urbana, IL 61801-3080 USA}

\begin{abstract}
The laser cooling and trapping of ultracold neutral dysprosium has been recently demonstrated using the broad, open 421-nm cycling transition.  Narrow-line magneto-optical trapping of Dy on longer wavelength transitions would enable the preparation of ultracold Dy samples suitable for loading optical dipole traps and subsequent evaporative cooling.  We have identified the closed 741-nm cycling transition as a candidate for the narrow-line cooling of Dy.  We present experimental data on the isotope shifts, the hyperfine constants $A$ and $B$, and the decay rate of the 741-nm transition.  In addition, we report a measurement of the 421-nm transition's linewidth, which agrees with previous measurements.  We summarize the laser cooling characteristics of these transitions as well as other narrow cycling transitions that may prove useful for cooling Dy. 

\end{abstract}
\date{\today}
\maketitle

\section{Introduction}

\begin{figure}[t]
\includegraphics[width=0.49\textwidth]{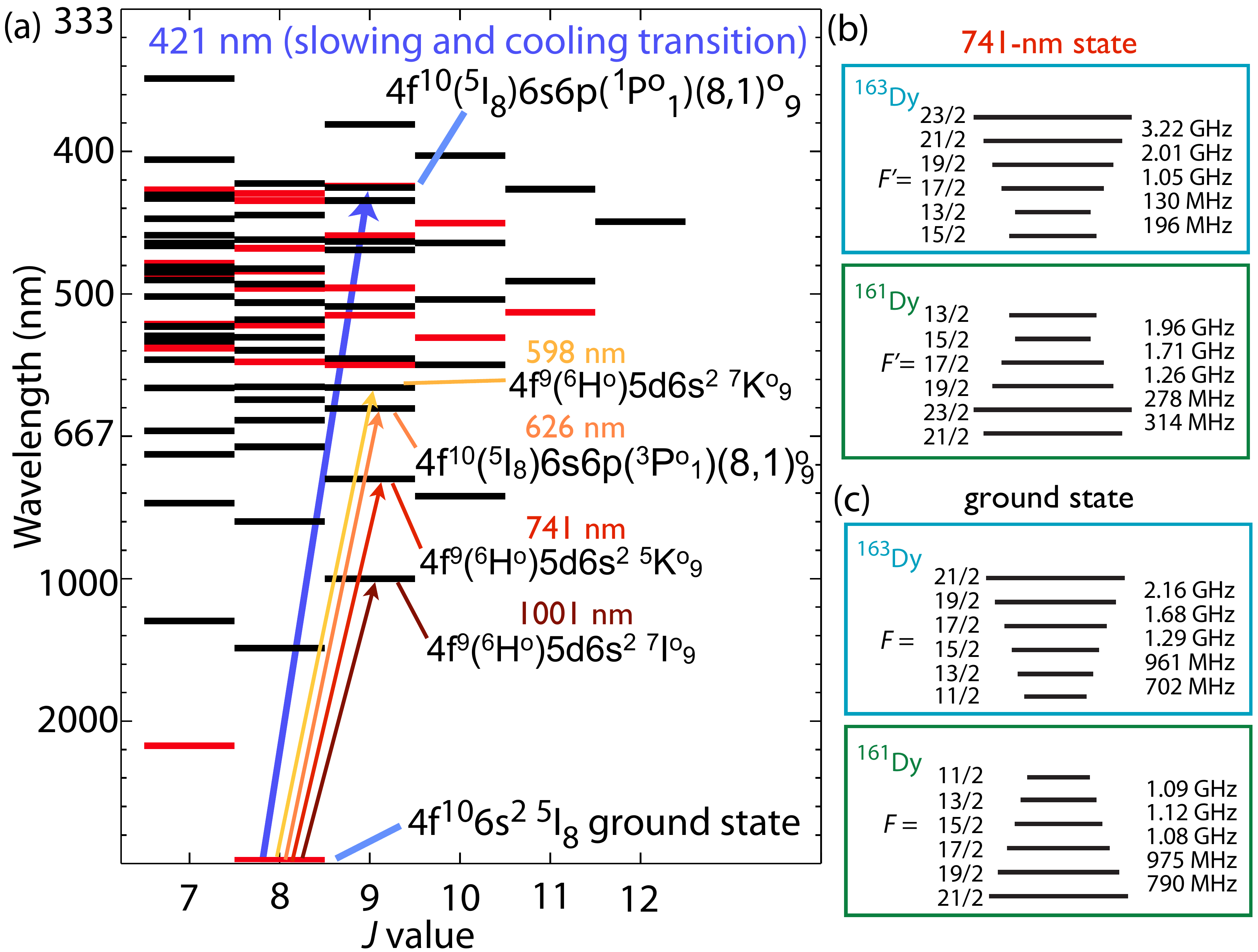}
\caption{(Color online)  (a) Dy energy level diagram with high $J$ values~\cite{Martin:1978}. The relevant laser cooling transitions between the even parity (red) ground state and the odd (black) excited states are marked with wavelengths and spectroscopic terms.  Dy's five high-abundance isotopes have nuclear spin $I=0$ for the bosons $^{164}$Dy, $^{162}$Dy, and $^{160}$Dy and $I=5/2$ for the fermions $^{163}$Dy and $^{161}$Dy.  (b) Fermion hyperfine structure in the 741-nm state (not to scale) determined from measurements in Sec.~\ref{hyperfine}.  (c) Fermion ground state hyperfine structure (not to scale)~\cite{Childs70}. $F=J+I$, where $J$ is the total electronic angular momentum.} 
\label{fig:dy_levels}
\end{figure}

Due to its extraordinarily large ground state magnetic dipole moment ($10$ Bohr magnetons), dysprosium is a prime candidate for the study of ultracold dipolar physics~\cite{PfauReview09,*Fregoso:2009,*Fregoso:2009b,*Machida10}.  The Dy atom belongs to the lanthanide (rare-earth) group and has ten $4f$ elections in an open shell submerged under closed $s$ shells.  Numerous combinations of valence electron couplings lead to a multitude of electronic energy levels.  Laser cooling Dy in a traditional manner would require an impracticable number of repumper lasers due to the large number of metastable states below 421 nm (see Fig.~\ref{fig:dy_levels}).  Consequently, preparation of cold Dy samples had been limited to the method of buffer gas cooling~\cite{Hancox:2004,*Newman2010}.  Recent progress in the laser cooling and trapping of Dy atoms in a repumper-less magneto-optical trap (MOT)~\cite{Lu2010} now allows the creation of large population samples at $>$100$\times$ colder temperatures.  Laser cooling and trapping Dy constitutes a new route toward achieving Bose and Fermi degeneracy in this most magnetic of atoms~\footnote{Fermionic Dy's magnetic moment is the largest of all elements, and bosonic Dy's is equal only to terbium's~\cite{Martin:1978}.}.   

However, further progress necessitates the optical trapping of Dy~\cite{MetcalfBook99} so that evaporative cooling may proceed without suffering trap losses arising from dipole-relaxation-induced inelastic collisions~\cite{Hensler:2003}.  A possible solution lies in the narrow-line-cooling of Dy in a manner similar to that demonstrated in the highly magnetic erbium system~\cite{Berglund:2008}.  A narrow-line Dy MOT could produce Dy temperatures in the $\mu$K-range.  Such ultracold samples would be readily confined in standard optical dipole traps (ODTs).

We focus here on the characteristics of the cycling transition at wavelength 741 nm.  We believe this transition to be a prime candidate for creating a narrow-line MOT in a similar manner as that demonstrated in Ref.~\cite{Berglund:2008} for Er.  Existing spectroscopic data~\cite{Martin:1978} are insufficient for implementing the 741-nm narrow-line MOT, and we present here the first measurements of this transition's hyperfine structure, isotope shifts, and lifetime.  Together these measurements provide a sufficient spectroscopic guide for attempting the 741-nm narrow-line cooling of Dy's bosonic and fermionic isotopes. 

Standard spectroscopic records~\cite{Martin:1978} had misrecorded by 21\% the linewidth of the 401-nm state used for Er laser cooling~\cite{Mcclelland:2006b}.  Linewidth verification of the analogous transition in Dy (at 421 nm) is therefore justified, and we present a linewidth measurement that is consistent with previous measurements.  Finally, we discuss the properties and relative merits of other optical transitions at 598 nm, 626 nm and 1001 nm that could also be used for laser cooling.

\section{741-nm transition}
We discuss in this section an optical transition at 741~nm that could prove important for creating narrow-line Dy MOTs. 
Although the broad, blue 421-nm transition is highly effective for cooling Dy atoms from an atomic beam and in a MOT~\cite{Youn2010a}, the transition's $\sim$1 mK Doppler limit is too high for directly loading an ODT.  Due to its narrow linewidth (unmeasured until now), the red 741-nm transition provides a means to Doppler cool with a very low temperature limit.  Our measurements indicate that this transition's recoil temperature would be higher than its Doppler-limited temperature: Observing novel laser cooling phenomena~\cite{Sr04} might be possible using this transition.

\subsection{Experimental apparatus for measuring isotope shifts and hyperfine structure}
The spectroscopic measurement for determining the isotope shifts and hyperfine structure employs a crossed excitation laser and atomic beam method~\cite{Demtroder}.  In a UHV chamber~\cite{Youn2010a}, thermal Dy atoms effuse from a high-temperature oven working at 1275  $^\circ$C.  The atoms are collimated by a long differential pumping tube which forms an atomic beam with a diameter of 5 mm and a diverging (half) angle of $\sim$0.02 rad. The beam enters a chamber with two pairs of optical viewports oriented orthogonally to the atomic beam.  
\begin{figure}[t]
\includegraphics[width=0.48\textwidth]{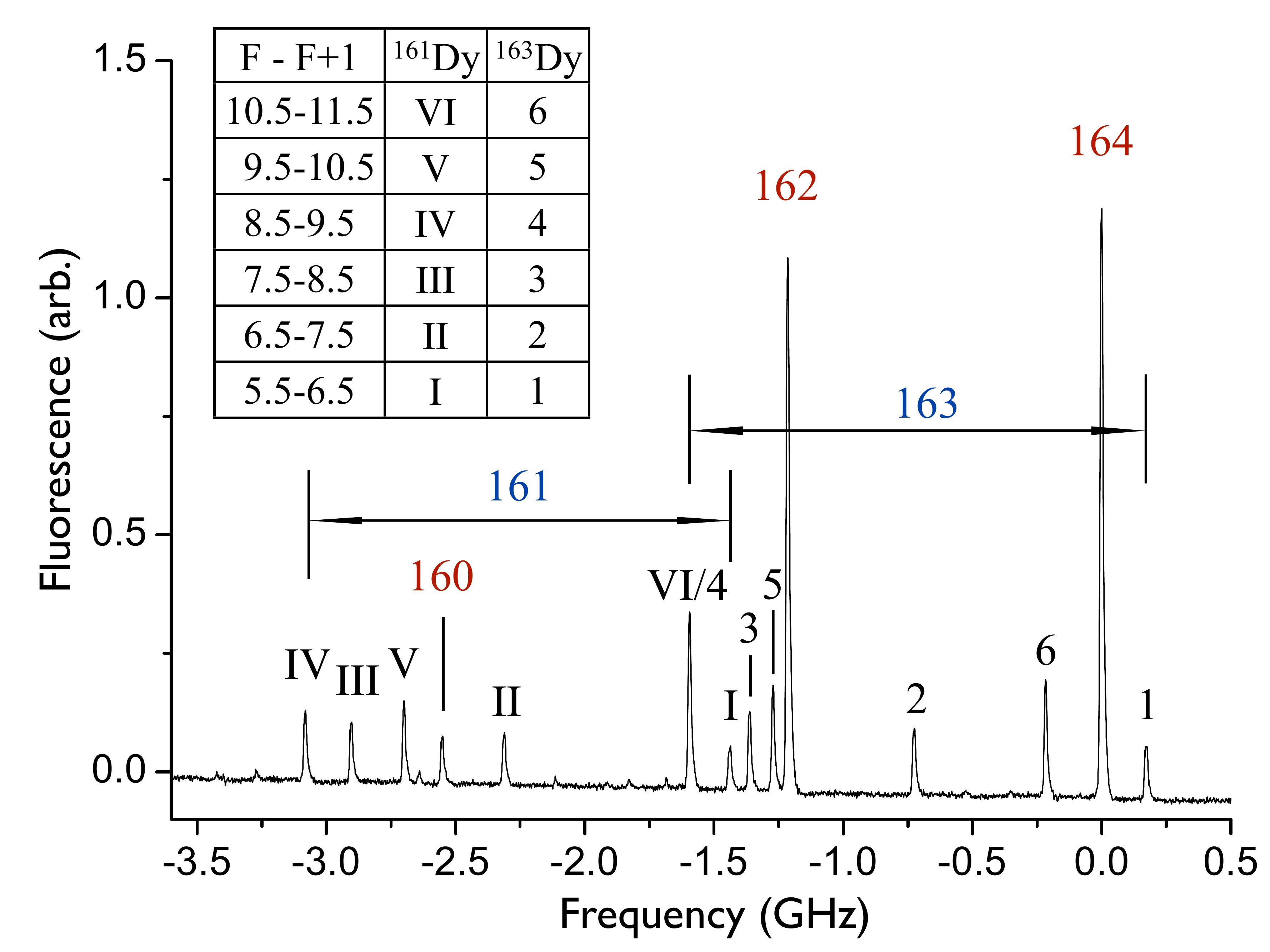}
\caption{(Color online)   741-nm line spectrum for the five most abundant Dy isotopes.  Bosonic isotope peaks are marked with mass numbers in red.  Hyperfine peaks of fermionic isotopes (blue) are identified by the markers defined in the inset table. The VIth peak of $^{161}$Dy and the 4th peak of $^{163}$Dy overlap.} 
\label{fig:741_spec}
\end{figure}

\begin{table} 

\caption{Values of hyperfine coefficients $A$ and $B$ (MHz) for the excited state ($e$) of the 741-nm transition in Dy including comparison with those of the ground state ($g$).}
\begin{ruledtabular} \label{ABtable}
\begin{tabular}{cccccc}

Coeff.   &  $^{163}$Dy$_\text{th}$\footnote{Ref.~\cite{Flambaum2010}}    &     $^{163}$Dy$_\text{expt}$   &       $^{161}$Dy$_\text{expt}$    &    $163_{e}/161_\text{e}$    &  $163_{g}/161_\text{g}$\footnote{Ref.~\cite{Childs70}}\\ 
\hline

$A$  &  142   &  142.91(3)  &   -102.09(2)    &   -1.3999(4)   &   -1.40026(4) \\
$B$  &   4000  &   4105(2)   &   3883(1)    & 1.0570(9)  &  1.05615(90)  \\ 

\end{tabular}
\end{ruledtabular}
\end{table}

 A 5 mW 741-nm laser beam [1/e$^{2}$ waist (radius) $\approx$ 2 mm] from an external cavity diode laser (ECDL) is directed through the UHV chamber via one pair of viewports. This laser has a mode-hop-free region of 20~GHz.  Atomic fluorescence on the 741-nm line is collected through an orthogonal viewport by a $2''$ AR-coated achromatic lens pair with a magnification of 0.4$\times$.  This forms an image on the detection area ($1$-mm diameter) of a femtowatt photodetector (DC gain: $1\times 10^{10}$ V/W, bandwidth $\geq$750~Hz).  The whole system is carefully shielded from stray light. 

%highly saturates the Dy atoms and effectively broadens the linewidth.

The direct output from the detector suffers from low signal-to-noise due to the oven's thermal radiation and the multiple scattering of 741-nm light from the windows and the chamber's inner walls.   An electronic bandpass filter (0.3 Hz to 3 kHz) with a DC amplification of 10 improves the signal-to-noise to a sufficient level without artificially broadening the Doppler-limited resonances.  With a laser scanning rate of $7$ GHz/s and a 15 MHz Doppler-broadening (measured in Sec.~\ref{421}), the $\sim$2 ms rise and fall time of the spectral peaks are slower than the detector's response time.   

To measure the full spectrum of isotope shifts and hyperfine states, we scan the ECDL using the piezoelectric transducer (PZT) that modulates its grating position.  However, the free scan of the ECDL suffers from the slight nonlinear scanning of the PZT versus drive voltage.  We reduce the nonlinearity by limiting the scan range and by scanning the PZT slowly to prevent inertial effects.  To calibrate the frequency scan, we couple the 741-nm light into a temperature-stabilized 750 MHz free-spectral range (FSR) confocal cavity.  Simultaneously recording the transmission of the confocal cavity with the fluorescence signal provides a frequency calibration as the ECDL is scanned.  The FSR of the cavity itself is measured via rf sidebands imprinted onto the cavity-coupled 741-nm light with a stable and calibrated rf frequency source.  To correct for the nonlinearity of the scan, a calibration is performed by fitting a polynomial to the cavity spectrum.  The maximum deviation throughout the scan due to a quadratic term is 3\% of the linear term (the cubic term is negligible); we corrected the nonlinear effect up to quadratic order. 

The calibrated spectrum after 512 averages is shown in Fig.~\ref{fig:741_spec}.  These data are sufficient to resolve and identify all $J\rightarrow J+1$ ($F\rightarrow F+1$) 741-nm transitions for the bosonic (fermionic) isotopes.  Optical pumping is a negligible effect in this spectrum since the $\sim$10 $\mu$s transit time is much shorter than the transition lifetime (see Sec.~\ref{741gamma}).

\subsection {Hyperfine structure}\label{hyperfine}
The position and ordering of $^{163}$Dy and $^{161}$Dy's hyperfine peaks in the spectrum are given by the ground and excited state's $A$ and $B$ coefficients~\footnote{The $A$ and $B$ coefficients are defined in, e.g., Ref.~\cite{MetcalfBook99}.}.    Identification of the isotope and hyperfine transition peaks are found with guidance from the calculations in Ref.~\cite{Flambaum2010}, and a least squares fitting routine extracts the experimental values of $A$ and $B$ (see Table~\ref{ABtable}).  Since this is a cycling $J\rightarrow J+1$ transition,  the strongest observed lines for the fermions are of the $F\rightarrow F+1$ type; the much weaker~\cite{Sobelman} $F \rightarrow F$ transitions are visible as the small, unlabeled peaks in Fig.~\ref{fig:741_spec}.  We note that for the excited states, the ratios of $A^{163}_{e}/A^{161}_{e} = -1.3999(4)$  and $B^{163}_{e}/B^{161}_{e} = 1.0570(9)$ are consistent with the analogous ratios for the ground state; there is no noticeable hyperfine anomaly for the 741-nm transition~\cite{Clark77}.

\subsection {Isotope shifts}

\begin{table}
\caption{Isotope shifts in the 741-nm and 457-nm transitions in Dy.}
\begin{ruledtabular}\label{isotopeshifts}
\begin{tabular}{ccc}
Isotope shifts & 741nm  & 457 nm\footnote{Ref.~\cite{Zaal80}} \\
\hline

$^{164}$Dy & 0   &  0\\
$^{163}$Dy & $-915(2)$ MHz & $660(3)$ MHz\\
$^{162}$Dy & $-1214(3)$ MHz & $971(2)$ MHz \\
$^{161}$Dy & $-2320(5)$ MHz & $1744(3)$ MHz\\
$^{160}$Dy & $-2552(5)$ MHz & $2020(3)$ MHz\\

\end{tabular}
\end{ruledtabular}
\end{table}

The measured isotope shifts (from $^{164}$Dy to $^{160}$Dy) for the 741-nm transition are listed in Table~\ref{isotopeshifts}, together with the isotope shifts for the 457-nm line.  The 457-nm line has a pure electronic configuration which makes it useful as a reference transition for creating the King plot~\cite{Budker} described below.  The fermionic isotope shifts  are derived from the center-of-gravity of the hyperfine peaks, the positions of which are extracted from hyperfine structure fits.  We could not obtain the isotope shift of $^{158}$Dy and $^{156}$Dy due to their low natural abundances and the weak strength of the 741-nm transition.  

We draw the King plot using the documented $4f^{10}6s^2({^5}I_8)\rightarrow 4f^{10}6s6p({^7}I_8)$ transition at 457 nm as the reference transition~\cite{Zaal80,Leefer:2009b}.  The isotope shift includes the contribution from the mass term, which is related to the change of isotope mass, and the field shift, which is due to the finite possibility of electrons being inside the nucleus.  Different electron configurations lead to different field shifts:  From the fit of the isotope shifts of the 741-nm transition plotted against  those values for 457 nm, the ratio of electronic field-shift parameter $E_{741}/E_{457}=-1.746(9)$ is determined based on the slope of the linear fit.  The minus sign indicates the very different nature of two transitions, i.e., one is of $4f\rightarrow5d$ type while the other is $6s\rightarrow6p$. 

The mass term includes the normal mass shift (NMS) and specific mass shift (SMS).  The NMS for the 457-nm and 741-nm transitions are $\delta\nu_{\text{nms}164-162}^{457}= 27$ MHz and $\delta\nu_{\text{nms}164-162}^{741}= 17$ MHz, respectively~\cite{Budker}. The SMS of the 457-nm transition is 7(8) MHz~\cite{Zaal80}, which allows us to calculate the SMS for the 741-nm line: $\delta\nu_{\text{sms}164-162}^{741}= 563(17)$ MHz, based on the intercept of the King plot. Such a large SMS is known to arise from the following effects:  $4f$ electrons deeply buried inside the electron core, strong electron correlations, and $4f$ electrons coupling to each other before coupling to the outer $6s$ electrons~\cite{Cowan73}.   
A transition of type $6s\rightarrow 6p$ leaves the inner electrons little changed, while $4f\rightarrow 5d$ transitions lead to large changes in the inner electron correlations. The large experimental value of the 741-nm transition SMS is consistent with the typical values for $4f\rightarrow 5d$ transitions~\cite{Jin01}.   

\begin{figure}[t]
\includegraphics[width=0.46\textwidth]{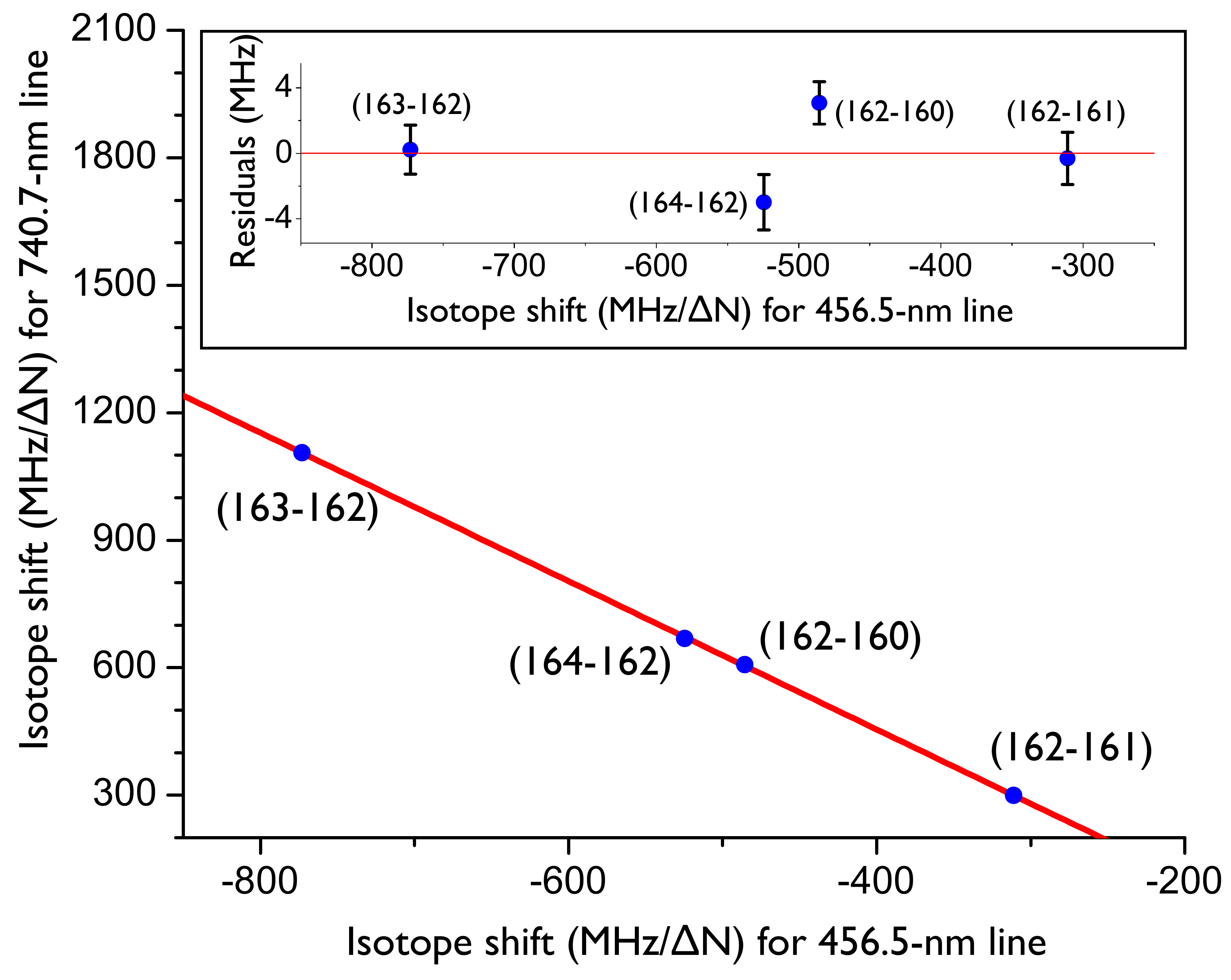}
\caption{(Color online) King plot of the isotope shifts in the 741-nm line versus isotope shifts in the 457-nm line.  $\Delta N$ is the mass number difference between isotope pairs.  Inset is the fit residual.} 
\label{fig:King plot}
\end{figure}

\subsection{Lifetime measurement}\label{741gamma}
A direct lifetime measurement based on the fluorescence decay observed with the crossed-beam method is not possible due to large transit time broadening relative to the natural decay time.  Therefore, we resort to measuring the fluorescence scattered from relatively static atoms, i.e., from the $\sim$1 mK atoms in the MOT and in the magnetostatic trap (MT).  We measure a consistent lifetime with both methods.

In the ``MOT'' method, we shine a retroreflected 5 mW 741-nm excitation beam of waist 3 mm onto a MOT generated on the 421-nm transition and record the decay of 741-nm scattered light after the 741-nm beam is extinguished.  While the 741-nm excitation beam is on, the system establishes a steady-state population distribution among the 421-nm, 741-nm, and ground states.   By switching on and off the 741-nm laser beam with a period of $250$ $\mu$s, the atoms initially shelved in the 741-nm  state will decay back to the ground state via the spontaneous emission of 741-nm photons.  The small $1:10^5$ branching ratio of the 421-nm transition, measured in previous work~\cite{Lu2010,Youn2010a}, means that the Dy atom is effectively a three-level system during the 250 $\mu$s decay measurement:  decay out of the three-level system from 421-nm state to the metastable states occurs on a much longer time scale, $>$2 ms.  Solutions to the optical Bloch equations~\cite{CohenT} for such a ``V'' system---two excited states coupled to a ground state via resonant 421-nm and 741-nm light---verify that the decay rate observed is equal to the decay rate of the bare 741-nm state.  

\begin{figure}[t]
\includegraphics[width=0.49\textwidth]{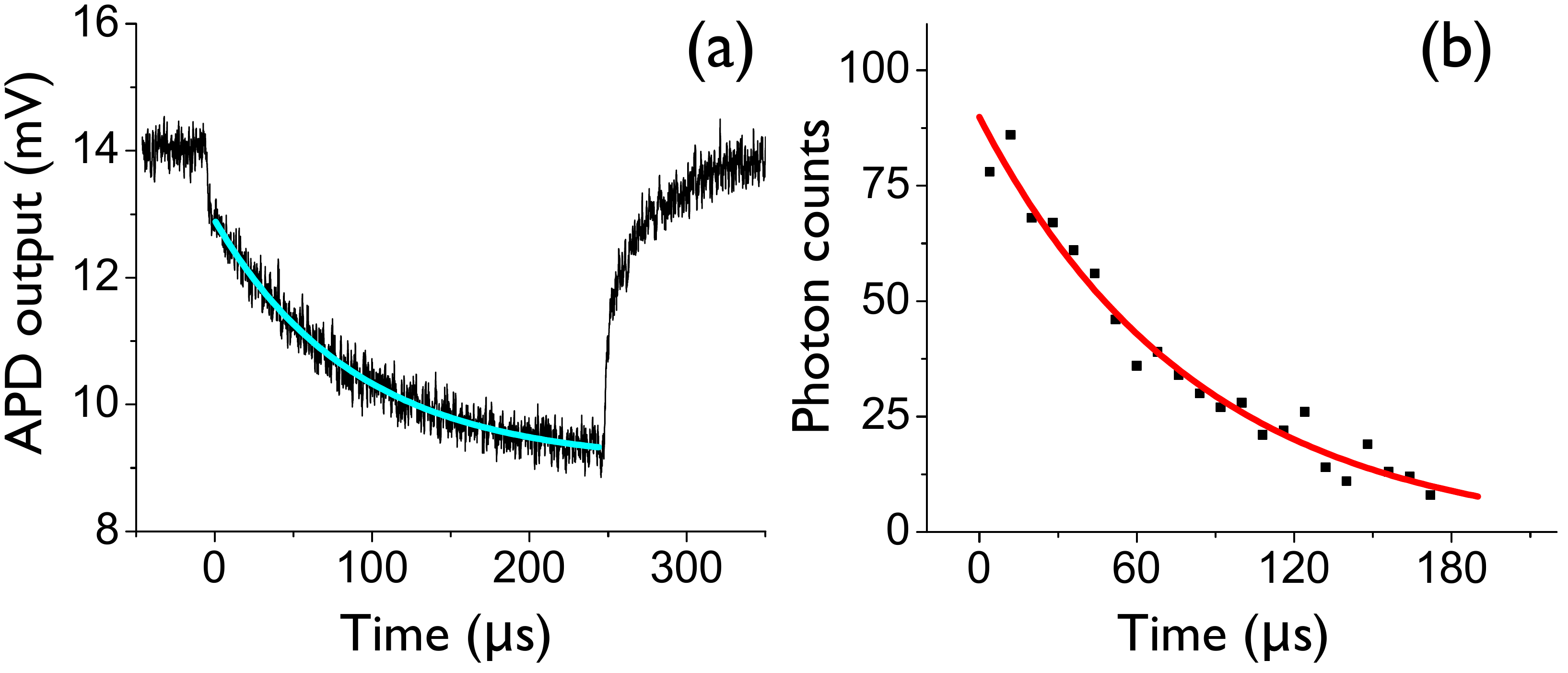}
\caption{(Color online)  Measured decay of 741-nm fluorescence. (a) MOT method:  Recorded 741-nm fluorescence signal from a Dy MOT by an APD averaged 10752 times.  The blue line is an exponential fit to the fluorescent decay of the 741-nm level. The 741-nm excitation laser is turned off (on) at $t=0$ ($t=250$) $\mu$s.  (b) MT method:  Photon counting record of scattered 741-nm light from a magnetic trap, averaged 27 times, after the 741-nm excitation beam is extinguished at $t=0$.  The red line is an exponential fit to the decay.} 
\label{fig:lifetime_741}
\end{figure}

An avalanche photodetector (APD) and collection lenses with a 741-nm narrow bandpass filter is used to detect the weak signal, which we average 10752 times to obtain the data in Fig.~\ref{fig:lifetime_741}(a).  Note that for neither this method nor the following one does the presence of 421-nm light or magnetic field gradients affect the natural decay rate measurement of the closed 741-nm transition.   

In the ``MT''  method, we extinguish the 421-nm MOT light and capture the atoms in the magnetic quadrupole field of the now-extinguished MOT.  We wait 5 s to allow the atoms to equilibrate in the MT~\cite{Lu2010} before shining onto the trap a resonant retroreflected 741-nm beam of 5 mW power and waist 3 mm.  A single photon counter with collection lenses and a 741-nm bandpass filter records the very weak flux of 741-nm photons from the MT after the 741-nm excitation beam is extinguished [see Fig.~\ref{fig:lifetime_741}(b)].  The long experimental run time necessary to measure a single decay limits the obtainable statistics. 

Single exponential fits to the data in both methods derive lifetimes that are consistent with each other (see Table~\ref{linewidth}). We note that the values are 4$\times$ longer than the theoretical value reported in Ref.~\cite{Flambaum2010}, and the measurement reported here may be used to refine Dy structure calculations~\cite{DzubaPC2010}. With such a narrow linewidth---the weighted combination of the lifetime is 89.3(8) $\mu$s, resulting in a linewidth of 1.78(2) kHz---narrow-line cooling on the 741-nm transition is technically challenging for red-detuned narrow-line MOTs, since the laser linewidth should be comparably narrow~\cite{Sr04}.  However, a blue-detuned narrow-line MOT, which relies on the atoms' large magnetic dipole and has been demonstrated with Er~\cite{Berglund:2008}, does not require a laser linewidth as narrow as the addressed atomic line;  a narrow-line blue-detuned MOT on the 741-nm line seems feasible.
\begin{table}[t]
\caption{Lifetime of the 741-nm excited state.}
\begin{ruledtabular}\label{linewidth}
\begin{tabular}{ccc}
MOT method& MT method & Theory\footnote{Ref.~\cite{Flambaum2010}} \\
\hline
$89.6(8)\,\rm{\mu s}$ & $84(14)\,\rm{\mu s}$& $21\,\rm{\mu s}$\\

\end{tabular}
\end{ruledtabular}
\end{table}

\section{421-nm transition}\label{421}

Quantitative understanding of the population, dynamics, and cooling mechanisms~\cite{Youn2010b} of the Dy MOT requires the accurate knowledge of the 421-nm transition's linewidth.  To ensure the use of the correct value of the 421-nm transition linewidth in laser cooling calculations, we remeasure this linewidth using the crossed-beam method described earlier, though with a 421-nm beam derived from a frequency-doubled Ti:Sapphire laser. 

To uniformly and stably scan the laser frequency, we employ the transfer cavity technique to lock the laser to a spectroscopic reference~\cite{Youn2010a}.  The optical transfer cavity is doubly resonant at 780 nm and 842 nm.  The cavity itself is stabilized by locking a 780-nm ECDL to a hyperfine transition of the 780-nm D2 line in Rb before locking a resonance of the cavity to the stabilized ECDL.   The Ti:Sapphire laser which generates the 842-nm beam is then locked to this cavity.  In order to scan the Ti:Sapphire laser's frequency while the cavity remains locked to Rb, an electro-optical modulator driven by a microwave source generates tunable GHz-frequency sidebands on the 842-nm laser beam.  By  locking the sideband to the cavity, the carrier frequency can be stably scanned via tuning the microwave source.  The 421-nm laser beam is obtained from a resonant LBO doubler.  In the experiment, the laser frequency is scanned 400 MHz with a period of 1 s to ensure that the bandwidth of the PIN photodetector does not artificially broaden the transition.  The fluorescence was collected via a pair of 2'' achromatic doublets mounted outside an AR-coated UHV viewport.  The electronic detector output was recorded on a fast digital oscilloscope and averaged 64 times.

\begin{figure}[t]
\includegraphics[width=0.45\textwidth]{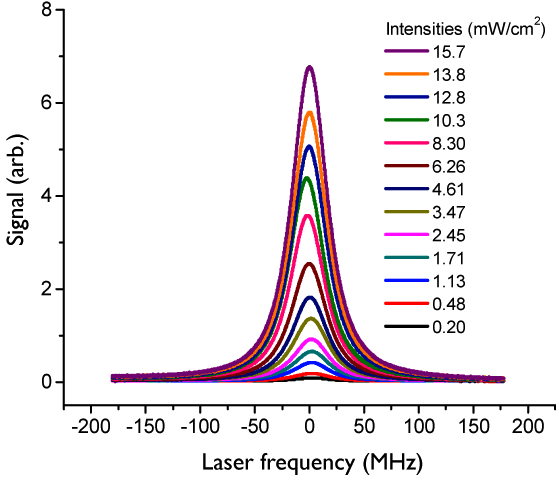}
\caption{(Color online) The photodetector signal as a function of 421-nm laser frequency, referenced to the line peak. The legend lists the laser intensities used in each measurement.  Each curve is averaged 64 times.} 
\label{fig:scan_421}
\end{figure}

The fluorescence versus frequency is shown in Fig.~\ref{fig:scan_421}.  The profile has the typical Voigt form due to the residual Doppler broadening of the atomic beam. Curves corresponding to different laser beam intensities possess differing linewidths due to power broadening.  A global Voigt fit allows a deconvolution of the Doppler width from the transition linewidth by assuming a single Gaussian Doppler width and by accounting for the power broadening from the laser.  The fitted value for the Doppler broadening is 14.8(6) MHz, which is consistent with the estimation of the residual Doppler broadening based on the geometry of the collimation tube and oven orifice.  

At low intensities, the power broadening is linear as a function of laser intensity.  A linear fit to the extracted linewidths provides the natural linewidth at the zero-intensity limit~\cite{Mcclelland:2006b}. The extrapolated value for the natural linewidth of the 421-nm transition is 31.9(8) MHz.  The uncertainty in the linewidth measurement arises from the error sources listed in Table~\ref{errors}.  Among the errors, the largest source is the laser frequency drift from the imperfect transfer cavity lock.  Unlike for Er~\cite{Mcclelland:2006b}, this measurement result is consistent with that listed in the standard tables, 33.1(17) MHz~\cite{Martin:1978,Lawler97}.

\section{Alternative laser cooling transitions}

Unlike the 421-nm line, the 598-nm, 626-nm, 741-nm and 1001-nm lines are closed cycling transitions; laser cooling on them would obviate the need for repumping lasers or magnetic confinement in metastable states. While the 421-nm transition has been used to form the first Dy MOT~\cite{Lu2010} and the 741-nm transition---easily generated by a stabiized ECDL---is a good candidate for narrow-line cooling, the other three transitions might also be useful for laser cooling and trapping. 

Table~\ref{Lines_sum} summarizes these five laser cooling transitions:  $g$ is the Land\'{e} factor of the excited state;  $\Gamma$ is the transition decay rate; the linewidth is $\Delta \nu = \Gamma/2\pi$; and the excited state lifetime is  $\tau = 1/\Gamma$.  From these values, we can calculate some quantities of importance to laser cooling and trapping~\cite{MetcalfBook99}.  The saturation intensity $I_\text{sat} \equiv \pi h c\Gamma/3\lambda^3$ is, e.g., an estimate of the required MOT laser power; the capture velocity $v_\text{cap} \equiv \Gamma\lambda/2\pi$ provides a measure of the velocity range within which atoms can be collected in a MOT; $T_\text{Doppler}  = \hbar\Gamma/2k_{B}$ is the Doppler cooling temperature limit; and $T_\text{recoil} = \hbar^2k^2/mk_B$ is the temperature limit due to photon recoils.

The 1001-nm transition was considered as a candidate for narrow-line cooling because this is an intercombination transition which typically possesses narrow linewidth.  We used the same experimental apparatus as in the 741-nm measurement---though a different ECDL laser---to find and measure the linewidth of the 1001-nm transition.  However, we did not detect the line.  Concurrently, theoretical calculations in Ref.~\cite{Flambaum2010} predicted the exceptionally small linewidth of $53$ Hz, which explains our inability to detect the line with our current apparatus.  This ultranarrow linewidth limits the transition's utility for a MOT, but along with the 741-nm line, the 1001-nm transition may be useful for resolved sideband cooling in an optical lattice~\cite{Katori03,Sterr07,Lev10}.  This cooling technique may provide an alternative method~\cite{Weiss02} to evaporative cooling for the production of degenerate Dy gases.
\begin{table}[t]
\caption{421-nm transition linewidth error budget.}
\begin{ruledtabular}\label{errors}
\begin{tabular}{cc}
Source & Uncertainty (MHz)\\
\hline
Extrapolation to zero intensity  & 0.13 \\
Drift of laser during scans\footnote{Scan nonlinearity is negligible in transfer cavity technique.}  &  0.7 \\
Laser linewidth   &  0.1 \\
Rise time of detector  & 0.4 \\

\end{tabular}
\end{ruledtabular}
\end{table}

The 626-nm transition has a intermediate linewidth of $135$ kHz, which could be used as the main laser cooling and trapping transition in a MOT while the atomic beam is Zeeman-slowed via the broad 421-nm transition. The benefit of such a combination~\cite{Yabuzaki99} lies in the lower MOT temperature, since the 626-nm transition's Doppler limit is only 3.2 $\mu$K.  A colder MOT facilitates subsequent ODT loading.  However, this combination requires the use of a narrow-linewidth dye-based laser to obtain 626-nm light.  We note that the large Land\'{e} $g$ factor difference between the excited state (1.29) and ground state (1.24) suggests that intra-MOT sub-Doppler cooling~\cite{Berglund:2007,Youn2010b} will not be as effective on this transition.

\begin{table*}[t]
\caption{Laser cooling parameters for five cycling transitions in Dy; see text for parameter definitions.}
\begin{ruledtabular}\label{Lines_sum}
\begin{tabular}{ccccccccc}
Line & $g$ & $\Gamma$ & $\Delta\nu$ & $\tau$ & $I_\text{sat}$ & $v_\text{cap}$ & $T_\text{Doppler}$ & $T_\text{recoil}$ \\
\hline

$421\ \rm{nm}$ &  $1.22$ & $2.00\times 10^8\ s^{-1}$ \footnote{Present work, 2.5\% uncertainty}& $31.9\ \rm{MHz}$ & $4.99\  \rm{ns}$ & $55.8\ \rm{mW/cm^{2}}$ & $13\ \rm{m/s}$ &$765\, \mu \rm{K}$ & $660\ \rm{nK}$\\

$598\ \rm{nm}$ &  $1.24$ & $7.7\times 10^4\ s^{-1} $ \footnote{Ref.~\cite{Flambaum2010}, theory} & $12\ \rm{kHz}$ & $13\ \rm{ \mu s}$ & $7.5\ \rm{\mu W/cm^{2}}$ & $7.3\ \rm{mm/s}$ &$294\, \rm{n K}$ & $327\ \rm{nK}$\\

$626\ \rm{nm}$ &  $1.29$ & $8.5\times 10^5\ s^{-1}$\footnote{Ref.~\cite{Martin:1978},   experiment 5\% uncertainty} & $135\ \rm{kHz}$ & $1.2\ \rm{\mu s}$ & $72\ \rm{\mu W/cm^{2}}$ & $8.5\ \rm{cm/s}$ &$3.2\, \rm{\mu K}$ & $298\ \rm{nK}$\\

$741\ \rm{nm}$ &  $1.23$ & $1.12\times 10^4\ s^{-1}$ \footnote{Present work, 1\% uncertainty} & $1.78\ \rm{kHz}$ & $89.3\ \rm{ \mu s}$ & $0.57\ \rm{\mu W/cm^{2}}$ & $1.3\ \rm{mm/s}$ &$42.7\, \rm{nK}$ & $213\ \rm{nK}$\\

$1001\ \rm{nm}$ &  $1.32$ & $3.3\times 10^2\ s^{-1}$\footnote{Ref.~\cite{Flambaum2010}, theory}  & $53\ \rm{Hz}$ & $3\ \rm{ ms}$ & $6.9\ \rm{nW/cm^{2}}$ & $0.05\ \rm{mm/s}$ &$1.3\, \rm{nK}$ & $116\ \rm{nK}$\\

\end{tabular}
\end{ruledtabular}
\end{table*}

The linewidth of the 598-nm transition has yet to be measured, but the calculated value~\cite{Flambaum2010} indicates a linewidth of $12$ kHz.  This narrow linewidth would be optimal for conventional narrow-line cooling as performed in, e.g., ultracold Sr experiments~\cite{Sr04}, but again, the line must be generated with a dye-based laser.  Its excited state Land\'{e} $g$ factor (1.24) is almost the same as its ground state's, which bodes well for effective intra-MOT sub-Doppler cooling.

\section{summary}

We measured the natural lifetime of the 741-nm line of Dy using a Dy MOT and magnetic trap; the weighted average is 89.3(8) $\mu$s.  We predict that this closed cycling transition will be useful for the formation of a narrow-line MOT, which could cool Dy to the ultracold temperatures necessary for loading an optical dipole trap.  The isotope shifts and hyperfine structure ($A$ and $B$ coefficients) were measured for all five high-abundance Dy isotopes, providing a spectral roadmap for the future narrow-line cooling of bosonic and fermionic Dy.  In addition, we verified the linewidth of the 421-nm transition in Dy to be 31.9(8) MHz, a precautionary measure taken since the standard tables had listed the analogous transition in Er to be in error by 21\%.  Finally, we tabulated---based on up-to-date linewidth information---the laser cooling properties of other attractive transitions in Dy.

\begin{acknowledgements}
We acknowledge support from the NSF (PHY08-47469), AFOSR (FA9550-09-1-0079), and the Army Research Office MURI 
award W911NF0910406. 
\end{acknowledgements}

%\bibliography{741bib}

\end{document}